\documentclass[aps,preprintnumbers,twocolumn]{revtex4}
\usepackage{amsfonts,amssymb,amsmath}            
\usepackage[]{graphics,graphicx,epsfig}            
\usepackage{amsthm}
\usepackage{subfigure}
\def\identity{\leavevmode\hbox{\small1\kern-3.8pt\normalsize1}}

\newcommand{\ket}[1]{\left | #1 \right\rangle}
\newcommand{\bra}[1]{\left \langle #1 \right |}

\newcommand{\braket}[2]{\left\langle #1|#2\right\rangle}
\newcommand{\proj}[1]{\ket{#1}\bra{#1}}
\begin{document}

\title{Local Cloning of Arbitrarily Entangled Multipartite States}
\date{\today}

\author{Alastair \surname{Kay}}
\affiliation{Centre for Quantum Computation,
             Centre for Mathematical Sciences,
             DAMTP,
             University of Cambridge,
             Wilberforce Road,
             Cambridge CB3 0WA, UK}
\author{Marie \surname{Ericsson}}
\affiliation{Centre for Quantum Computation,
             Centre for Mathematical Sciences,
             DAMTP,
             University of Cambridge,
             Wilberforce Road,
             Cambridge CB3 0WA, UK}

\begin{abstract}
We examine the perfect cloning of non-local, orthogonal states
with only local operations and classical communication. We
provide a complete characterisation of the states that can
be cloned under these restrictions, and their relation to distinguishability.
We also consider the case of catalytic cloning,
which we show provides no enhancement to the set of clonable states.
\end{abstract}

\maketitle

Cloning and entanglement are key features of quantum information
theory. The no-cloning theorem \cite{noclone,noclone2}, for example,
has implications for the security of quantum cryptosystems
\cite{bb84}. Similarly, entanglement is typically viewed as a
resource that we can use to enhance various processes, such as the security of
communication \cite{e91}. Significant effort has been put into
developing a theory of entanglement, which is elucidated by considering the
restriction to local operations and classical
communication (LOCC). However, little is known about the
intersection of these two ideas, such as the resource requirements for
cloning a set of orthogonal states that are distributed
between two or more parties. Previous works have shown that cloning 
maximally entangled states (MESs) is, in principle, possible
\cite{cloners, Hayashi:2004}, but have not broached the subject of less entangled
states.

Cloning is a well-defined operation, where you start
with a quantum state which is one of a set of states,
$\{\ket{\psi_i}\}$, but you do not know which. The task is to create
a second copy. It is well known \cite{nielsen} that this can only be
done perfectly for orthogonal states, i.e.
$$
\braket{\psi_i}{\psi_j}=\delta_{ij}.
$$
The key is that the party (or parties) trying to perform the cloning
operation must be able to distinguish between the possible states
perfectly.

The situation that we wish to tackle here is when an unknown
bipartite pure state from the set $\{\ket{\psi_i}\}$ is distributed between Alice and
Bob. Whether cloning can be accomplished depends on what resources they share. The case
where Alice and Bob can only perform LOCC, but do not share any
other resources has already been studied \cite{Horodecki:2004a},
where it was shown that, provided the states are distillable, they
cannot be cloned.

If Alice and Bob share three MESs as a resource, Bob can always
teleport his state to Alice, who can distinguish between the complete
set of states and then they can recreate two copies of the
state on the two remaining MESs. If they share two MESs then a pair of
orthogonal states can always be locally distinguished \cite{Walgate:2000} with projective measurements and two
copies created from the MESs. For higher dimensional systems the size of the set is often
larger than just two states \cite{Walgate:2002}.

\begin{figure}
\begin{center}
\includegraphics[width=0.3\textwidth]{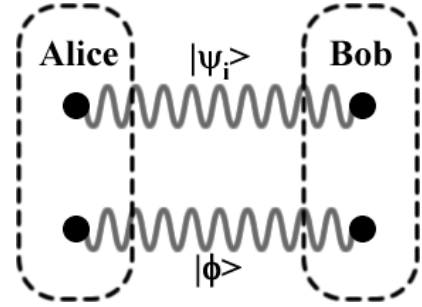}
\caption{Alice and Bob
share the state to be cloned, $\ket{\psi_i}$, and a maximally
entangled state, $\ket{\phi}$. Each of the four sub-systems is $d$-dimensional.} \label{fig:protocol}
\end{center}
\end{figure}

The case that we are interested in is when Alice and Bob share only
a single MES as a resource, as depicted in Fig. \ref{fig:protocol}. This places the question at a fine
division where there exists enough entanglement to, in principle, be
able to perform the cloning operation but not sufficient that Alice and Bob
can afford to make measurements on their state (which would destroy
the entanglement). Previous investigations \cite{cloners,
Hayashi:2004} have concentrated on restricting the $\ket{\psi_i}$ to
also being MESs. In particular, it was demonstrated in \cite{cloners}
that if $\ket{\psi_0}$ is maximally entangled, the whole set of states
$\{\ket{\psi_i}\}$ must also be maximally entangled. It was
further proven that the maximal size of the set of bipartite MESs that
can be cloned is equal to the size of dimension of Alice's (or
Bob's) system, $d$ \cite{Hayashi:2004}. In that paper, the necessary
and sufficient conditions for cloning these states were derived.

In this paper, we present an alternative interpretation of the
results in \cite{Hayashi:2004} and extend the proof to non-maximally
entangled states. As such, we entirely answer the question of which
states can be cloned. We will then consider the extension to
multiparties, and address the question of whether a catalyst can
contribute anything to the cloning process.

We would first like to provide some motivation for which bipartite states
can be cloned perfectly by only LOCC and an additional MES. It is
reasonable to think that Alice and Bob should be able to distinguish
between the states with measurements and only a single round of
two-way classical communication (by distinguishing we mean
that both Alice and Bob know which state they shared). This is, in
essence, because the MES that is shared can be used to replace this
single round of measurements and communication \cite{popescu:2000}.
For example, it is shown in \cite{Walgate:2000} that any two
orthogonal two-qubit states (with extension to higher dimensional
and multipartite states) can be represented in the form
\begin{eqnarray}
\ket{\psi_0}&=&a\ket{00}+b\ket{1}\ket{u}    \nonumber\\
\ket{\psi_1}&=&c\ket{01}+d\ket{1}\ket{u_{\!\perp}}  \nonumber
\end{eqnarray}
where $\braket{u}{u_\perp}=0$, i.e. we can get to states of this
form by local rotations only.  To distinguish between these states,
Alice and Bob, in general, have to perform two rounds of communication,
i.e. Alice makes a measurement in the $\{\ket{0},\ket{1}\}$ basis
and communicates the result to Bob, who can choose to measure in the
$\{\ket{0},\ket{1}\}$ basis if Alice measured $\ket{0}$ or in the
$\{\ket{u},\ket{u_\perp}\}$ basis if Alice measured $\ket{1}$.
Then Bob has to communicate his result to Alice so that they both
know which state they had. In the special case where $\ket{u}=\ket{1}$,
Alice and Bob are able to distinguish the state by both performing
measurements in the $\{\ket{0},\ket{1}\}$ basis and then they simultaneously
send the results to each other. We therefore expect to find that
these states can be cloned under the restrictions of LOCC, but not pairs of states such as
\begin{eqnarray}
\ket{\psi_0}&=&a\ket{00}+b\ket{11}  \nonumber\\
\ket{\psi_1}&=&b\ket{00}-a\ket{11}  \nonumber
\end{eqnarray}
(unless $a=1/\sqrt{2}$). We also expect that to be able to perform LOCC
cloning, the states should have the same Schmidt
coefficients (same entanglement). This is because we need to perform a
POVM on the MES in order to reduce its entanglement.
This POVM will be independent of which state is being cloned, and hence
we expect the coefficients to come out the same.

Before considering how to clone non-MESs we would like to return to
the previously discussed results on cloning MESs. In
\cite{Hayashi:2004} it is shown that for MESs of dimension $d$, any
set $\{\ket{\psi_i}\}$, of up to $d$ states, that can be locally
copied obeys the relation
$\ket{\psi_j}=(U_j\otimes\identity)\ket{\psi_0}$, where
$$
U_j=\sum_{k=0}^{d-1}\omega^{jk}\proj{k}
$$
and $\omega$ is the $d^{th}$ root of unity. Let us now reinterpret
this in terms of distinguishability. By performing the basis change
$\ket{k}\rightarrow\sum_m\omega^{mk}\ket{m}$, we observe that $U_j$
is equivalent, up to local unitaries, to the permutation operation
$P_{-j}$, where
$$P_i=\sum_{j=0}^{d-1}\ket{j+i \text{ mod }
d}\bra{j}.$$ Hence, measuring in this adjusted basis, Alice and Bob
can distinguish these states with a single round of communication.
This provides the alternative interpretation that if, and only if, Alice
and Bob can distinguish between the entire set of MESs to be cloned
with a single round of measurements and classical communication,
they can be cloned perfectly.

Now, let us move on to the question of LOCC cloning of non-MESs with
one MES as a resource. We consider, first, the case when Alice and Bob make
projective measurements on the unknown state $\ket{\psi_i}$ to
determine $i$ \cite{Walgate:2000}, and then use the single MES
together with a separable state to create two copies of
$\ket{\psi_i}$. To check if this is possible, we just have to verify
whether the majorization condition holds \cite{majorization}. If the MES
that we are using as an additional resource is represented by 
$\ket{\phi}=\frac{1}{\sqrt{d}}\sum_{l=0}^{d-1}\ket{l}\ket{l}$, then the
initial state $\ket{\phi}\ket{00}$ has
$d$ non-zero Schmidt coefficients (all $1/d$). As a result, the
target state $\ket{\psi_i}\ket{\psi_i}$ must only have $d$ non-zero
Schmidt coefficients out of a possible $d^2$, i.e. $\ket{\psi_i}$
must have an effective dimension $\sqrt{d}$. We reinterpret this by
saying that, in order to clone a set of states $\{\ket{\psi_i}\}$ of
dimension $d$ by this method, we {\em always} require a MES of
dimension at least $d^2$, or two MESs of dimension $d$. This means
that we are not able to clone with projective measurements on
the unknown state.

Instead of performing projective measurements, we shall now consider
a more general protocol. The first state that we want to clone can
always be written in its Schmidt basis,
$$
\ket{\psi_0}=\sum_{i=0}^{d-1}\alpha_i\ket{i}\ket{i}
$$
We want to consider the most general possible set of operations that
can result in the creation of perfect clones. To do this, we shall
imagine Alice and Bob performing a whole series of POVMs, each in
response to all the previous results of measurements by either
party. For perfect cloning, we must achieve cloning for every single
sequence of measurement results. Let us therefore pick one possible
sequence of outcomes. The result of these measurements can thus be
described by operators $M\otimes N$ applied by Alice and Bob
respectively. If we are to achieve cloning, the following must hold,
$$
M\otimes N\ket{\psi_0}\ket{\phi}=\sqrt{p_0}\ket{\psi_0}\ket{\psi_0}.
$$
This equation allows $M$ to be written in terms of $N$. We choose a representation for $N$,
$$N=U_\lambda \left(\sum_{i=0}^{2d-1}\beta_i\proj{i}\right)U_\eta,$$
where $U_\lambda$ and $U_\eta$ are arbitrary unitaries over two qudits.
In fact,
we find that we can always write
\begin{equation}
M=\left(\sum_{i=0}^{d-1}\alpha_i\proj{i}\right)^{\otimes 2} N'
\left[\left(\sum_{i=0}^{d-1}\frac{1}{\alpha_i}\proj{i}\right)\otimes\identity\right]
\label{eqn:rep}
\end{equation}
where
$$
N'=U_\lambda^*\left(\sum_{i=0}^{2d-1}\frac{1}{\beta_i}\proj{i}\right)U_\eta^*
$$
up to some constant of proportionality. The two qudits on which $M$
is defined are Alice's qudits from the unknown state $\ket{\psi_i}$
and the MES, $\ket{\phi}$. $N'\otimes N$  (or more precisely $N$,
since $N'$ is completely determined by $N$) then has to be picked such that
the other $\ket{\psi_i}$s are also cloned. However, we can derive simple
conditions from understanding the above equation. In particular, the
right-most term in eqn. (\ref{eqn:rep}) converts $\ket{\psi_0}$ into a MES. From
there, $N$ performs cloning of this MES, and then the entanglement
of both MESs is reduced with the left-most terms. Recalling the previous results about MESs, we conclude
that for $N$ to perform the cloning,
$\left[\left(\sum_{j=0}^{d-1}\frac{1}{\alpha_j}\proj{j}\right)\otimes
\identity\right]\ket{\psi_i}$ must, up to normalisation, be
maximally entangled and distinguishable by a single round of
classical communication, i.e.
$$
\ket{\psi_i}=\sum_{j=0}^{d-1}\alpha_j\ket{j}\ket{j+i}=\left(\identity\otimes
P_i\right)\ket{\psi_0}.
$$

Having proved the necessity of the conditions for cloning, we shall
now demonstrate sufficiency by providing a protocol that clones
these states using a single MES as a resource. The set of (up to) $d$
states $\{\ket{\psi_i}\}$ that we want to clone can be written as
\begin{eqnarray}
\ket{\psi_i}&=&\sum_{j=0}^{d-1}\alpha_j\ket{j}P_i\ket{j}. \nonumber
\end{eqnarray}
When both Alice and Bob perform the operation
\begin{equation}
U=\sum_{m=0}^{d-1}\proj{m}\otimes P_m,
\label{eqn:unitary}
\end{equation}
the maximally entangled state is converted to
$$
\ket{\phi_i}=\frac{1}{\sqrt{d}}\sum_{l=0}^{d-1}\ket{l}P_i\ket{l}
$$
as a result of the property $P_j\otimes P_j\ket{\phi_i}=\ket{\phi_i}$.
The $\ket{\phi_i}$s are each a different, orthogonal, state for each of the unknown states $\ket{\psi_i}$.
Note
that, in the case of $d=2$, $U$ is the controlled-NOT gate. We then
have to convert the MES into a less entangled
state. This is achieved by Alice applying the measurement operators
\begin{eqnarray}
M_0&=&\sum_{j=0}^{d-1}\alpha_j\proj{j}        \label{eqn:meas}\\
M_k&=&P_kM_0P_k^\dagger.        \nonumber
\end{eqnarray}
If Alice gets the result $k$, then both Alice and Bob apply the
correction $P_k^\dagger\equiv P_{-k}$. This performs the required
conversion $\ket{\phi_i}\rightarrow\ket{\psi_i}$, for all possible
measurement results $k$, hence completing the cloning protocol. This protocol is easily linked to eqn. (\ref{eqn:rep}), because the cloning of MESs can be achieved with a unitary, so $N=N'=U$ (eqn. (\ref{eqn:unitary})). Further, the two POVMs that Alice performs on $\ket{\psi_i}$ in eqn. (\ref{eqn:rep}) commute with this unitary, and hence cancel. This just leaves the measurements $M_i$ (eqn. (\ref{eqn:meas})) applied to the MES.

Such a protocol extends to multiple parties in a straightforward
way. For example, the tripartite case has a set of $d^2$ states that
can be cloned,
$$
\ket{\psi_{ij}}=\sum_{k=0}^{d-1}\alpha_k\ket{k}P_i\ket{k}P_j\ket{k},
$$
provided the three parties also share a state of the form
$$
\ket{\phi}=\frac{1}{\sqrt{d}}\sum_{i=0}^{d-1}\ket{i}\ket{i}\ket{i}.
$$
In this case, they follow exactly the same protocol as before,
where the third party, Charlie, does exactly the same as Bob
does. To justify that these are the only states that can be cloned,
we borrow an argument from \cite{Walgate:2000}. The authors describe
how to distinguish between multipartite states by reducing the
number of parties involved. Consider, for example, the states
\begin{eqnarray}
\ket{\psi_0}&=&\ket{0}_{A}\ket{\Gamma _{0}}_{BC}+\cdots +\ket{l}_{A'}
\ket{\Gamma _{l}}_{BC} \nonumber\\
\ket{\psi_1}&=&\ket{0}_{A}\ket{\Gamma _{0}^{\perp
}}_{BC}+\cdots+\ket{l}_{A}\ket{\Gamma _{l}^
{\perp}}_{BC} \nonumber
\end{eqnarray}
where the states $\ket{\Gamma}$ are not normalised. We can
immediately see that Alice and Bob will need to be able to clone the
set of bipartite states $\{\ket{\Gamma_i},\ket{\Gamma_i^\perp}\}$,
which simply reduces to the previous condition.

We would also like to tackle the question of catalytic cloning. In
this situation, not only are we provided with a MES on which to
create the clone, but some other entangled state
which can be used in the protocol, but must be returned unchanged at
the end of the cloning process.
$$
\ket{\psi_i}\ket{\phi}\ket{C}\rightarrow\ket{\psi_i}\ket{\psi_i}\ket{C}
$$
This state, $\ket{C}$, acts as a catalyst, in much the same way as
conversion between some states can only occur with the help of a
catalyst \cite{catalyst}. The states
$\ket{\psi_i}$ have Schmidt coefficients $\alpha^i_j$, and the
catalyst has Schmidt coefficients $\beta_j$. Repeating the previous
argument shows that $\{\alpha_i^1\beta_j\}=\{\alpha_i^2\beta_j\}$.
It is clear that this isn't true except in the cases that we could
already clone. To see this, consider the smallest values of
$\alpha_i^1$ and $\beta_j$ - if $\min\alpha_i^1\neq\min\alpha_i^2$,
then there are values that we cannot account for. We can then
progress through the hierarchy of next-largest $\alpha_i^1$, until
we find that the sets $\{\alpha_i^1\}$ and $\{\alpha_i^2\}$ must be
the same. Hence, a catalyst cannot provide any enhancement in the cloning process.

In conclusion, we have completely characterised what can be achieved
with the local cloning of non-local states. The set of clonable
states is very restrictive - they must be locally distinguishable
with a single round of two-way communication, and they must have the
same entanglement. We have demonstrated a protocol that clones all
these states. In addition, we have ruled out the possibility of
catalytic cloning.

We would like to thank Roger Colbeck and Daniel Oi for useful
discussions. In addition, ME thanks Lucien Hardy for introductory conversations.
Adrian Kent, Artur Ekert and Daniel Oi are acknowledged as being
part of the bet which prompted this paper. AK is supported
by EPSRC. ME acknowledges financial support from the Swedish Research
Council.


\begin{thebibliography}{13}
\expandafter\ifx\csname natexlab\endcsname\relax\def\natexlab#1{#1}\fi
\expandafter\ifx\csname bibnamefont\endcsname\relax
  \def\bibnamefont#1{#1}\fi
\expandafter\ifx\csname bibfnamefont\endcsname\relax
  \def\bibfnamefont#1{#1}\fi
\expandafter\ifx\csname citenamefont\endcsname\relax
  \def\citenamefont#1{#1}\fi
\expandafter\ifx\csname url\endcsname\relax
  \def\url#1{\texttt{#1}}\fi
\expandafter\ifx\csname urlprefix\endcsname\relax\def\urlprefix{URL }\fi
\providecommand{\bibinfo}[2]{#2}
\providecommand{\eprint}[2][]{\url{#2}}

\bibitem[{\citenamefont{Wootters and Zurek}(1982)}]{noclone}
\bibinfo{author}{\bibfnamefont{W.~K.} \bibnamefont{Wootters}} \bibnamefont{and}
  \bibinfo{author}{\bibfnamefont{W.~H.} \bibnamefont{Zurek}},
  \bibinfo{journal}{Nature} \textbf{\bibinfo{volume}{299}},
  \bibinfo{pages}{802} (\bibinfo{year}{1982}).

\bibitem[{\citenamefont{D.Dieks}(1982)}]{noclone2}
\bibinfo{author}{\bibnamefont{D.Dieks}}, \bibinfo{journal}{Phys. Lett. A}
  \textbf{\bibinfo{volume}{92}}, \bibinfo{pages}{271} (\bibinfo{year}{1982}).

\bibitem[{\citenamefont{Bennet and Brassard}(1984)}]{bb84}
\bibinfo{author}{\bibfnamefont{C.~H.} \bibnamefont{Bennet}} \bibnamefont{and}
  \bibinfo{author}{\bibfnamefont{G.}~\bibnamefont{Brassard}},
  \bibinfo{journal}{Proceedings of IEEE International Conference on Computers,
  Systems and Signal Processing} p. \bibinfo{pages}{175}
  (\bibinfo{year}{1984}).

\bibitem[{\citenamefont{Ekert}(1991)}]{e91}
\bibinfo{author}{\bibfnamefont{A.~K.} \bibnamefont{Ekert}},
  \bibinfo{journal}{Phys. Rev. Lett.} \textbf{\bibinfo{volume}{67}},
  \bibinfo{pages}{661} (\bibinfo{year}{1991}).

\bibitem[{\citenamefont{Anselmi et~al.}(2004)\citenamefont{Anselmi, Chefles,
  and Plenio}}]{cloners}
\bibinfo{author}{\bibfnamefont{F.}~\bibnamefont{Anselmi}},
  \bibinfo{author}{\bibfnamefont{A.}~\bibnamefont{Chefles}}, \bibnamefont{and}
  \bibinfo{author}{\bibfnamefont{M.}~\bibnamefont{Plenio}},
  \bibinfo{journal}{NJP} \textbf{\bibinfo{volume}{6}}, \bibinfo{pages}{164}
  (\bibinfo{year}{2004}), \bibinfo{note}{quant-ph/0407168}.

\bibitem[{\citenamefont{Owari and Hayashi}(2004)}]{Hayashi:2004}
\bibinfo{author}{\bibfnamefont{M.}~\bibnamefont{Owari}} \bibnamefont{and}
  \bibinfo{author}{\bibfnamefont{M.}~\bibnamefont{Hayashi}}
  (\bibinfo{year}{2004}), \bibinfo{note}{quant-ph/0411143}.

\bibitem[{\citenamefont{Nielsen and Chuang}(2000)}]{nielsen}
\bibinfo{author}{\bibfnamefont{M.~A.} \bibnamefont{Nielsen}} \bibnamefont{and}
  \bibinfo{author}{\bibfnamefont{I.~L.} \bibnamefont{Chuang}},
  \emph{\bibinfo{title}{Quantum Computation and Quantum Information}}
  (\bibinfo{publisher}{CUP}, \bibinfo{year}{2000}), \bibinfo{edition}{5th} ed.

\bibitem[{\citenamefont{Horodecki et~al.}(2004)\citenamefont{Horodecki, Sen,
  and Sen}}]{Horodecki:2004a}
\bibinfo{author}{\bibfnamefont{M.}~\bibnamefont{Horodecki}},
  \bibinfo{author}{\bibfnamefont{A.}~\bibnamefont{Sen}}, \bibnamefont{and}
  \bibinfo{author}{\bibfnamefont{U.}~\bibnamefont{Sen}},
  \bibinfo{journal}{Phys. Rev. A} \textbf{\bibinfo{volume}{70}},
  \bibinfo{pages}{052326} (\bibinfo{year}{2004}).

\bibitem[{\citenamefont{Walgate et~al.}(2000)\citenamefont{Walgate, Short,
  Hardy, and Vedral}}]{Walgate:2000}
\bibinfo{author}{\bibfnamefont{J.}~\bibnamefont{Walgate}},
  \bibinfo{author}{\bibfnamefont{A.~J.} \bibnamefont{Short}},
  \bibinfo{author}{\bibfnamefont{L.}~\bibnamefont{Hardy}}, \bibnamefont{and}
  \bibinfo{author}{\bibfnamefont{V.}~\bibnamefont{Vedral}},
  \bibinfo{journal}{Phys. Rev. Lett.} \textbf{\bibinfo{volume}{85}},
  \bibinfo{pages}{4972} (\bibinfo{year}{2000}).

\bibitem[{\citenamefont{Walgate and Hardy}(2002)}]{Walgate:2002}
\bibinfo{author}{\bibfnamefont{J.}~\bibnamefont{Walgate}} \bibnamefont{and}
  \bibinfo{author}{\bibfnamefont{L.}~\bibnamefont{Hardy}},
  \bibinfo{journal}{Phys. Rev. Lett.} \textbf{\bibinfo{volume}{89}},
  \bibinfo{pages}{147901} (\bibinfo{year}{2002}).

\bibitem[{\citenamefont{Collins et~al.}(2001)\citenamefont{Collins, Linden, and
  Popescu}}]{popescu:2000}
\bibinfo{author}{\bibfnamefont{D.}~\bibnamefont{Collins}},
  \bibinfo{author}{\bibfnamefont{N.}~\bibnamefont{Linden}}, \bibnamefont{and}
  \bibinfo{author}{\bibfnamefont{S.}~\bibnamefont{Popescu}},
  \bibinfo{journal}{Phys. Rev. A} \textbf{\bibinfo{volume}{64}},
  \bibinfo{pages}{032302} (\bibinfo{year}{2001}).

\bibitem[{\citenamefont{Nielsen}(1999)}]{majorization}
\bibinfo{author}{\bibfnamefont{M.~A.} \bibnamefont{Nielsen}},
  \bibinfo{journal}{Phys. Rev. Lett.} \textbf{\bibinfo{volume}{83}},
  \bibinfo{pages}{436} (\bibinfo{year}{1999}).

\bibitem[{\citenamefont{Jonathan and Plenio}(1999)}]{catalyst}
\bibinfo{author}{\bibfnamefont{D.}~\bibnamefont{Jonathan}} \bibnamefont{and}
  \bibinfo{author}{\bibfnamefont{M.~B.} \bibnamefont{Plenio}},
  \bibinfo{journal}{Phys. Rev. Lett.} \textbf{\bibinfo{volume}{83}},
  \bibinfo{pages}{3566} (\bibinfo{year}{1999}).

\end{thebibliography}
\end{document}